\documentclass[aps,prb,twocolumn,groupedaddress]{revtex4}
\usepackage{graphicx}
\usepackage{amssymb}
\usepackage{wasysym}
\usepackage{pifont}

\newcommand{\beq}{\begin{equation}}
\newcommand{\eeq}{\end{equation}}

\begin{document}

\title{Current-Induced Effective Magnetic Fields in Co/Cu/Co Nanopillars}

\author{M. A. Zimmler, B. \"{O}zyilmaz, W. Chen and A. D. Kent}
\affiliation{Department of Physics, New York University, New York, NY 10003, USA}

\author{J. Z. Sun, M. J. Rooks and R. H. Koch}
\affiliation{IBM T. J. Watson Research Center, P.O. Box 218, Yorktown Heights, NY 10598, USA}

\date{\today}

\begin{abstract}
We present a method to measure the effective field contribution to spin-transfer-induced interactions between the magnetic layers in a trilayer nanostructure, which enables spin-current effects to be distinguished from the usual charge-current-induced magnetic fields. This technique is demonstrated on submicron Co/Cu/Co nanopillars. The hysteresis loop of one of the magnetic layers in the trilayer is measured as a function of current while the direction of magnetization of the other layer is kept fixed, first in one direction and then in the opposite direction. These measurements show a current-dependent shift of the hysteresis loop which, based on the symmetry of the magnetic response, we associate with spin-transfer. The observed loop-shift with applied current at room temperature is reduced in measurements at 4.2 K. We interprete these results both in terms of a spin-current dependent effective activation barrier for magnetization reversal and a spin-current dependent effective magnetic field. From data at 4.2 K we estimate the magnitude of the spin-transfer induced effective field to be $\sim 1.5 \times 10^{-7} $ Oe cm$^2$/A, about a factor of 5 less than the spin-transfer torque.
\end{abstract}

\pacs{}

\maketitle

\section{Introduction}
A spin-polarized current may interact with the magnetic moment of a thin-film nanomagnet causing the reversal of magnetization (switching) or other magnetic excitations such as spin waves. These spin-transfer effects are of fundamental importance in understanding spin-transport at the nanoscale and are also of technological relevance in the development of a new class of spin-electronic devices such as current-switched magnetic memories and current-tunable microwave sources \cite{Kiselev2003,Rippard2004}.

The general form of the spin-transfer torque, introduced in 1996 by Slonczewski \cite{Slonczewski1996} and Berger \cite{Berger1996}, has found strong support in experiments \cite{Tsoi1998,Sun1999,Myers1999,Wegrowe1999,Katine2000,Tsoi2000,Grollier2001,Ozyilmaz2003}. Alternate models have been proposed, however, which consider an additional ``effective field'' interaction between the magnetic moments in a multilayered structure. Heide \emph{et al.} \cite{Heide2001a,Heide2001b} proposed a model in which the effective field arises as a result of the \emph{longitudinal} component of the spin accumulation, which magnetically ``couples'' the layers to produce switching. An alternate diffusive model was proposed by Zhang \emph{et al.} \cite{Zhang2002}, in which the interaction between the local moments and the \emph{transverse} component of the spin accumulation leads to the Slonczewski torque term, $ a_J \, \hat{m} \times ( \hat{m} \times \hat{m_P} ) $, and an additional effective field term of the form $ b_J \, \hat{m} \times \hat{m_P} $, where $ \hat{m} $ and $ \hat{m_P} $ are unit vectors in the directions of the magnetization of the ``free'' and ``fixed'' magnetic layers, respectively. The coefficients $ a_J $ and $ b_J $ both depend linearly on current and the ratio of their magnitudes depends on the thickness of the free layer and the decay length of the transverse component of the spin accumulation. In fact, any torque on the magnetization of the free layer can be decomposed into these two mutually perpendicular components in the plane orthogonal to $ \hat{m} $. Brataas \emph{et al.} \cite{Brataas2001} also considered a model with an effective field. In this model, the effective field is related to the imaginary part of the mixing conductance $ G^{\uparrow \downarrow} $ while the spin torque is related to its real part. For metallic ferromagnets, $ \mathrm{Re} G^{\uparrow \downarrow}$ is about ten times  $\mathrm{Im} G^{\uparrow \downarrow} $ \cite{Xia2002}, a similar conclusion to that reached by Slonczewski \cite{Slonczewski2002}. A related model by Waintal \emph{et al.} \cite{Waintal} and a model by Stiles and Zangwill \emph{et al.} \cite{Stiles2002} found no effective field type interaction.
\begin{figure}[t]
\includegraphics[width=0.48\textwidth]{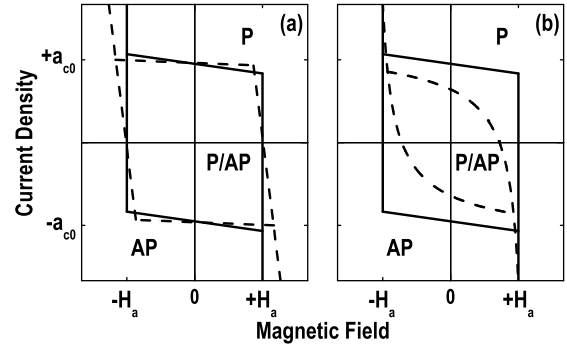}
\caption{Solid lines: Form of the zero-temperature stability diagram in the Slonczewski torque model for a single domain magnet with uniaxial anisotropy. Dashed lines: (a) Stability diagram with an additional effective field term of the form $ b_J \, \hat{m} \times \hat{m_P} $, $ b_J < 0 $. (b) Finite-temperature stability diagram in the Slonczewski torque model. In these schematic phase diagrams we have omitted the regions in which there are precessional states for clarity \cite{Myers2002,Krivorotov}.}
\label{EffFieldPD}
\end{figure}

Within this framework, in the zero-temperature, monodomain approximation, the motion of the free layer magnetization under the influence of a spin-current is described by the Landau-Lifshitz-Gilbert (LLG) equation with the Slonczewski torque term and an effective field term,
\begin{eqnarray}
\lefteqn{\frac{d\hat{m}}{\gamma dt} = -\hat{m} \times \left[ \vec{H}_{\mathrm{eff}} - \alpha \, \frac{d\hat{m}}{\gamma dt} \right]} \\
& & \mbox{} - a_J \, \hat{m} \times ( \hat{m} \times \hat{m}_P ) + b_J \, \hat{m} \times \hat{m}_P. \nonumber
\label{eom}
\end{eqnarray}
$ \vec{H}_{\mathrm{eff}} $ is the effective field, $ \vec{H}_{\mathrm{eff}} = ( H_{\mathrm{app}} + H_{\mathrm a} \, m_x) \hat{x} - 4 \pi M m_z \hat{z} $, where $ H_{\mathrm{app}} $, the applied external field, and $ H_{\mathrm a} $, the in-plane uniaxial anisotropy field, are both in the $ \hat{x} $ direction, $ \hat{z} $ is the film normal and $ M  $ is the free layer magnetization. $ \gamma $ is the gyromagnetic ratio and $ \alpha $ is the Gilbert damping parameter ($ \alpha \ll 1 $). The fixed layer magnetization $\hat{m}_P$ is also taken to be in the $ \hat{x} $ direction. The coefficient in the Slonczewski term, $ a_J $, depends on the current density, the spin polarization $ P $ and the angle between the free and fixed magnetic layers $ \hat{m} \cdot \hat{m}_P $. Figure \ref{EffFieldPD}(a) shows a schematic phase diagram obtained by numerically integrating Eq. (\ref{eom}) with the effective field term (dashed lines) and without it (solid lines). This effective field term leads to a linear current-dependent shift of the field axis, $ H_{\mathrm{app}} \rightarrow H_{\mathrm{app}} - b_J $. The vertical boundaries which correspond to the external magnetic field overcoming the in-plane anisotropy field $ H_{\mathrm a} $ would be rotated clockwise or counterclockwise about $ J = 0 $, depending on the algebraic sign of $ b_J $. The spin torque term, however, does not effect this boundary at zero temperature. Below its threshold value, $ |a_J| < a_{c0} = \alpha 2 \pi M $, the spin current  modulates the damping \cite{Li2003}, but the magnitude of the damping does not change the switching field of the nanomagnet. Switching still occurs at the critical field when the metastable state becomes unstable.

Finite temperatures may be modeled with the addition of a Langevin random field $ \vec{H}_L $ to the effective field $\vec{H}_{\mathrm{eff}} $ in Eq. (\ref{eom}) \cite{Brown1963,Grinstein2003}. This results in a finite probability for thermally activated switching with an activation barrier to magnetization reversal that depends linearly on the current \cite{Li2004,Myers2002,Koch2004}. Figure \ref{EffFieldPD}(b) shows a schematic phase diagram where finite-temperature effects have been included. As can be seen from this figure and as we show below, at finite temperature the spin-torque term also leads to a shift of the midpoint of the hysteresis loop. However, if an effective field as that proposed by Zhang \emph{et al.} were to be significant, its effects should be noticeable at low temperatures where the influence of thermal fluctuations becomes negligible. Experiments have ruled out the possibility of \emph{only} a spin-transfer induced (STI) effective field with no spin-transfer torque interaction, but not the possibility of both mechanisms concurrently \cite{Myers2002,Grollier2003,Koch2004}. Further, there have been no attempts to directly measure the magnitude of the STI effective field interaction.

It is important to note that charge-current induced (CCI) magnetic fields may also produce a shift of the hysteresis loop that is linear in the applied current. For example, if the contact resistance is comparable to that of the pillar, the current may flow asymmetrically producing a net magnetic field bias on the nanomagnet. Thus, if we wish to measure a STI effective field, either one arising from a term of the form $ b_J \, \hat{m} \times \hat{m}_P $ or one associated with the Slonczewski torque term and finite temperature, we must identify the contribution from CCI magnetic fields.\begin{figure}[t]
\includegraphics[width=0.48\textwidth]{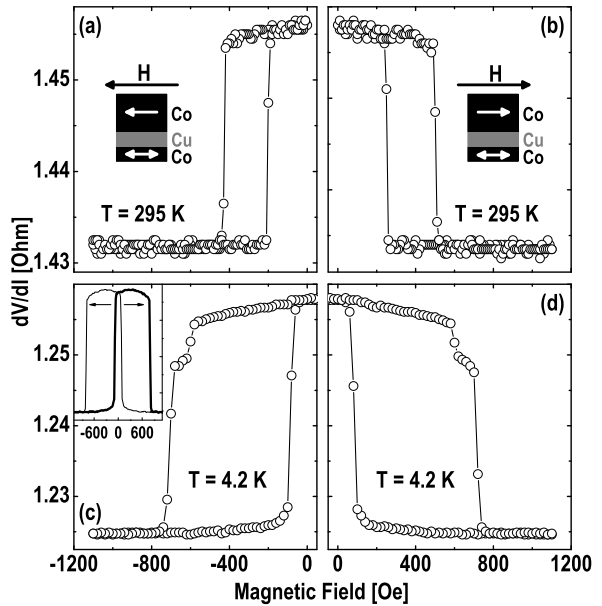}
\caption{Differential resistance of a nanopillar spin valve device as a function of magnetic field measured at $ T = 295 $ K (a,b) and $ T = 4.2 $ K (c,d), for the fixed Co layer oriented in the $ +\hat{x} $ (b,d) and $ -\hat{x} $ (a,c) direction. The insets in (a,b) show the corresponding magnetic configurations of the layers. (c) Inset: Full-sweep differential resistance of the trilayer structure as a function of magnetic field measured at $ T = 4.2 $ K. Note that in this figure both layers eventually switch in the direction of the applied magnetic field, while in (a,b,c,d) only the free layer switches its orientation. The vertical scale is the same as in (c,d).}
\label{MR}
\end{figure}

Here we describe a method to measure the effective field contribution of the spin-transfer mechanism that distinguishes the effect of a charge current from that of a spin current on the switching characteristics of the thin magnetic layer within a Co/Cu/Co trilayer nanopillar. Measurements at room temperature exhibit a linear shift of the center of the hysteresis loop of the thin magnetic layer with applied current, which is reduced in measurements at 4.2 K. We interprete these results in terms of both a Slonczewski torque term, including thermal fluctuations of the magnetization, and a spin-current-dependent effective magnetic field. Based on measurements at 4.2 K for which the influence of thermal fluctuations on the switching fields is negligible, we estimate the magnitude of a STI effective field of the form $ b_J \, \hat{m} \times \hat{m}_P $ and compare this to the magnitude of the spin-torque term, $ a_J \, \hat{m} \times ( \hat{m} \times \hat{m_P} ) $.

\section{Experimental method}
We study samples fabricated by thermal and electron beam evaporation through a nano-stencil mask process \cite{Sun2002,Sun2003}, with the stack sequence $|$3 nm Co$|$10 nm Cu$|$12 nm Co$|$300 nm Cu$|$10 nm Pt$|$. The thick ``fixed'' Co layer sets up a spin-polarization of the current, which then acts on the thin ``free'' Co layer causing the reversal of the orientation of its magnetization for large enough current densities ($ J \approx 10^8 $ A/cm$^2$). Although the results we present below correspond to one sample of lateral size 50 nm $\times$ 100 nm, similar results were obtained on several other samples with different lateral size dimensions. Transport measurements were conducted at 295 K and 4.2 K in a four-point measurement geometry, where we measured the differential resistance $ dV/dI $ (and dc voltage simultaneously) by means of a phase sensitive lock-in technique with a 200 $\mu$A modulation current, at $ f = 750 $ Hz, added to a dc bias current. The resistance per square of the top and bottom leads was found to be 0.65 $\Omega$, to be compared to the nanopillar resistance of 1.4 $\Omega$. The external magnetic field was swept at a rate of 12.8 Oe/s for measurements at $ T = 295 $ K and 20.3 Oe/s for measurements at $ T = 4.2 $ K. We define positive currents so that electrons flow from the fixed to the free Co layer.

The orientation of the STI effective field is determined by the direction of the magnetization of the fixed Co layer. In particular, an effective field of the form $ b_J \, \hat{m} \times \hat{m}_P $ would tend to align the magnetization of the free layer $ \hat{m} $ in the direction of that of the fixed layer $ \hat{m}_P $. Reversing the orientation of the fixed layer, \emph{i.e.} $ \hat{m}_P \rightarrow -\hat{m}_P $, would simply change the sign of the effective field. On the other hand, CCI magnetic fields do not depend on whether the magnetic layers are aligned in the $ +\hat{x} $ or $ -\hat{x} $ direction. If $ H_s $ denotes the STI effective fields and $ H_c $ the CCI magnetic field, this symmetry may be written as
\begin{center}
\begin{tabular}{c}
$ H_s(\hat{m}_P) = BJ \rightarrow H_s(-\hat{m}_P) = -BJ $ \\
$ H_c(\hat{m}_P) = CJ \rightarrow H_s(-\hat{m}_P) = CJ, $ \\
\end{tabular}
\end{center}
where $ H_s $ and $ H_c $ have both been expressed as linear functions of the current density. We thus see that in one magnetic configuration the CCI magnetic fields reinforce the STI fields while in the other configuration they oppose the spin-transfer effects. Figure \ref{MR} shows these two magnetic configurations schematically together with the differential resistance versus magnetic field, with zero bias current, obtained for each configuration, at 295 K (Figs. \ref{MR}(a) and (b)) and at 4.2 K (Figs. \ref{MR}(c) and (d)). Note the marked change in the width of the hysteresis loop from 295 K to 4.2 K. The field drives the free layer moment hysteretically between a low-resistance state parallel (P) to the fixed Co layer and a higher-resistance antiparallel (AP) state. The magnetic coupling between the fixed and free Co layers (either exchange or dipolar) causes the midpoint of the hysteresis loop to be shifted from $ H = 0 $.
\begin{figure}[t]
\includegraphics[width=0.48\textwidth]{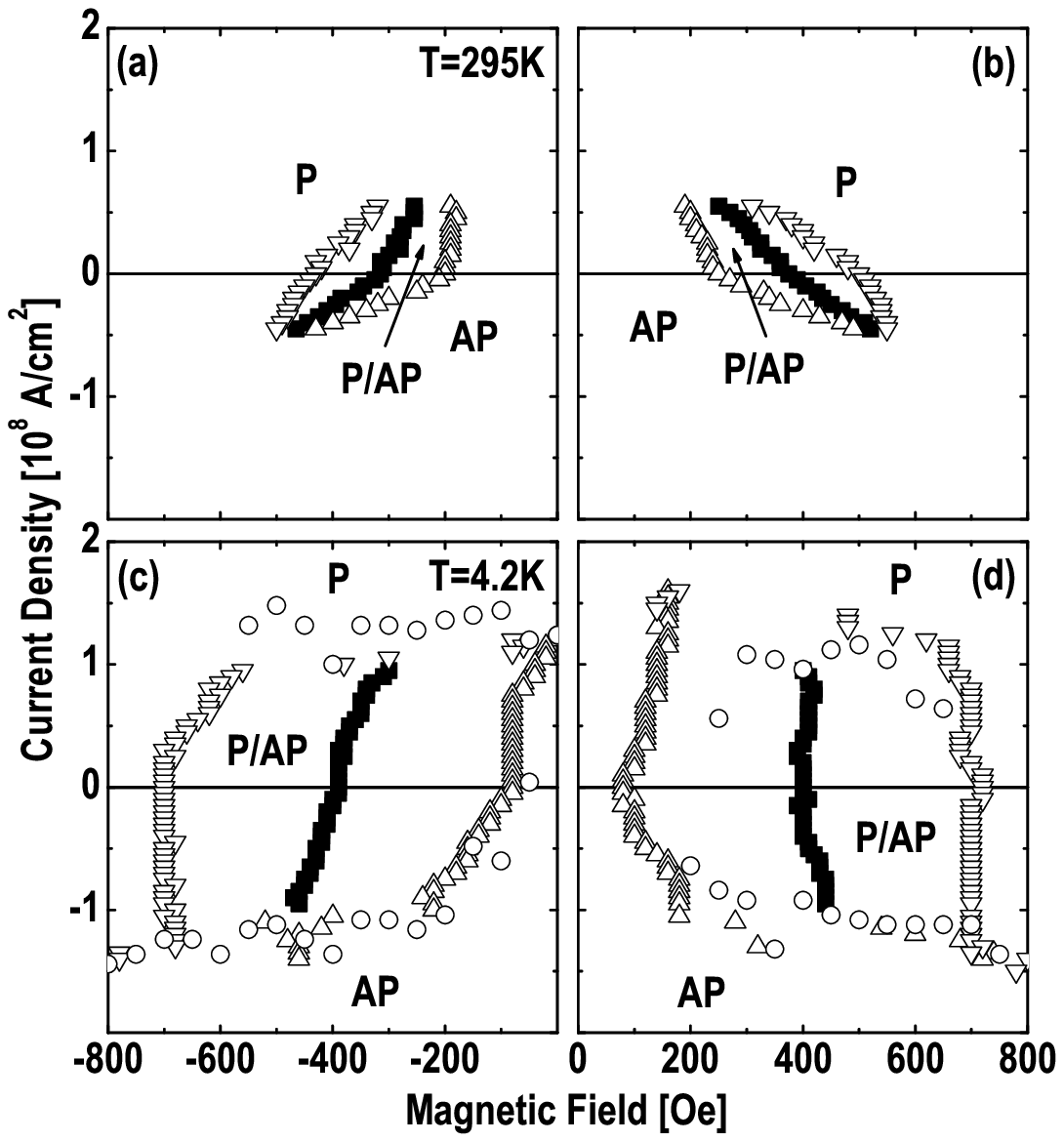}
\caption{Phase diagrams of magnetic states for the free Co layer obtained from measurements of $ dV/dI $ as a function of $ H $ at fixed $ I $ ($ \bigtriangleup $, $ H_{\mathrm{P} \rightarrow \mathrm{AP}} $;  $ \bigtriangledown $, $ H_{\mathrm{AP} \rightarrow \mathrm{P}} $) and as a function of $ I $ at fixed $ H $ ($ \Circle $), measured at $ T = 295 $ K (a,b) and  $ T = 4.2 $ K (c,d), for the fixed Co layer oriented in the $ +\hat{x} $ (b,d) and $ -\hat{x} $ (a,c) direction. Solid squares ($ \blacksquare $) show the observed effective bias fields $H_J^+$ (b,d) and $H_J^-$ (a,c).}
\label{PD}
\end{figure}
Carrying out magnetic field hysteresis loops for the free layer for \emph{both} orientations of the fixed layer, we are thus able to distinguish STI effective fields from CCI magnetic fields. The combined current-induced effective field on the nanomagnet can be decomposed into a charge-current and a spin-current component. By simply defining the observed total effective field as
\beq
H_J^{\pm} = \frac{H_{\mathrm{AP} \rightarrow \mathrm{P}}^{\pm} + H_{\mathrm{P} \rightarrow \mathrm{AP}}^{\pm}}{2},
\label{EffField}
\eeq
where the $ + $ or $ - $ corresponds to the effective field with the fixed layer in the $ +\hat{x} $ or $ -\hat{x} $ direction, it is easily seen from the preceeding discussion that the effective field contribution which only depends on the orientation of the fixed layer (STI) is given by
\beq
H_s = \frac{H_J^+ - H_J^-}{2}
\label{STEF}
\eeq
and the contribution which is independent of the magnetic configuration (CCI) by
\beq
H_c = \frac{H_J^+ + H_J^-}{2}.
\label{CCIMF}
\eeq
It is important to note that hysteresis loops must be performed on the free layer \emph{only}. If we were to ramp the magnetic field so as to first switch the free layer and then the fixed layer (Fig. \ref{MR}(c), inset), the effective field would be ``reset'' as we changed the orientation of the fixed layer and the free layer AP $ \rightarrow $ P field transition would not be measured.
\section{Results}
In Fig. \ref{PD}, we construct phase diagrams based on measurements of $ dV/dI $ as a function of $ H $ at fixed $ I $, measured at $ T = 295 $ K (Figs. \ref{PD}(a) and (b)) and at $ T = 4.2 $ K (Figs. \ref{PD}(c) and (d)), for the fixed Co layer along the $ +\hat{x} $ (Figs. \ref{PD}(b) and (d)) and $ -\hat{x} $ (Figs. \ref{PD}(a) and (c)) direction. The figure shows at what values of $ H $ and $ I $ the different static states are stable or bistable. Note that, unlike previous measurements \cite{Urazhdin2003}, our phase diagrams do not include the switching of the fixed layer (See the inset in Fig. \ref{MR}(c)). Some of the switching boundaries are thus distinct from those obtained in field sweeps in which both fixed and free layers switch, in particular, the free layer AP $ \rightarrow $ P transition. Qualitatively, however, there are similarities between the two types of measurements, such as with respect to the effects of temperature. At $ T = 4.2 $ K the field stability boundaries are almost independent of the applied current. The corresponding boundaries at $ T = 295 $ K, on the other hand, vary significantly with current. Also, while at $ T = 295 $ K the current dependence is  clearly asymmetric, at $ T = 4.2 $ K the asymmetry is small.

\begin{figure}[t]
\includegraphics[width=0.48\textwidth]{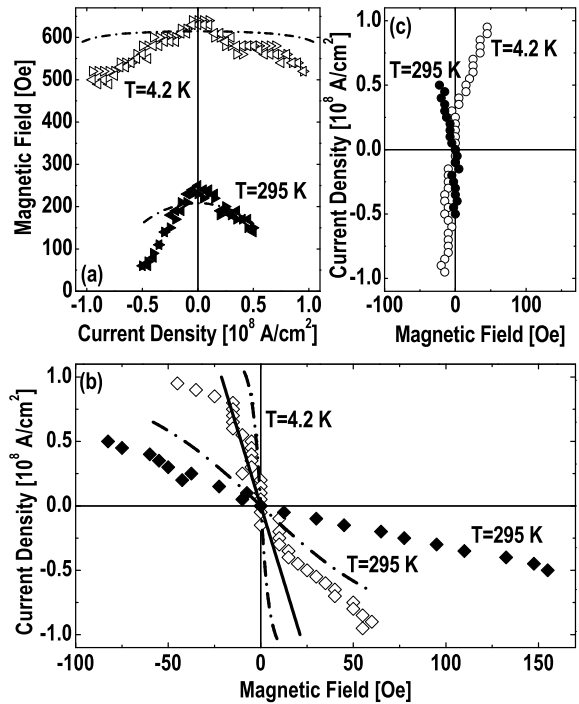}
\caption{Solid symbols denote measurements at $ T = 295 $ K and open symbols denote measurements at $ T = 4.2 $ K. (a) Hysteresis loop width as a function of current density for the fixed Co layer oriented in the $ +\hat{x} $ ($ \vartriangleright $, $ \blacktriangleright $) and $ -\hat{x} $ ($ \vartriangleleft $, $ \blacktriangleleft $) direction. The dashed line for $ T = 295 $ K was obtained by fitting the data near $ J = 0 $ to Eq. (\ref{hwidth}), with $ A = (ak_BT/U_0)^{1/\beta} $ as the fitting parameter. The dashed line for $ T = 4.2 $ K is a plot of Eq. (\ref{hwidth}) with the value of $ U_0 $ obtained from the $ T = 295 $ K fit, $ \beta = 3/2 $ and $ a = 20 $. (b) STI effective field extracted from the data in Fig. \ref{PD} using Eq. (\ref{STEF}). The dashed lines are plots of Eq. (\ref{hbias}) with the parameter $ U_0 $ obtained from the fit to the $ T = 295 $ K hysteresis loop width data, $ \beta = 3/2 $ and $ a = 20 $. The solid line is a linear fit to the STI effective field, near $ J = 0 $, at $ T = 4.2 $ K. (c) CCI magnetic field obtained from Fig. \ref{PD} using Eq. (\ref{CCIMF}).}
\label{Results}
\end{figure}

Figure \ref{Results} shows the different contributions to the total effective field extracted from the data in Fig. \ref{PD}. In Figs. \ref{Results}(b) and (c) we have used Eqs. (\ref{STEF}) and (\ref{CCIMF}), respectively, after substracting the magnetic coupling field (determined from the shift from $ H = 0 $ of the midpoint of the hysteresis loop for $ J = 0 $), to deduce the contributions from the fields that depend on magnetic configuration (STI) and those that do not (CCI). Figure \ref{Results}(a) shows the current dependence of the width of the hysteresis loop for $ T = 295 $ K and $ T = 4.2 $ K.

From the data in Fig. \ref{Results}(b) it is clear that the effective field $ H_s $ is strongly dependent on temperature. While CCI magnetic fields are small both at room temperature and at 4.2 K (except for a clear effect for $ |J|> 0.3 \times 10^8 $ A/cm$^2$, which corresponds to the appearance of magnetic domains), the strong current dependence of the effective field at room temperature is reduced at 4.2 K. This data suggests that thermal fluctuations play an important role in determining the switching boundaries. We thus consider the influence of finite temperature before estimating the magnitude of the STI effective field interaction in these structures.

\section{Finite Temperature Effects}
Finite temperature has been shown to have an important effect on the spin-transfer induced switching thresholds \cite{Li2004,Myers2002,Koch2004}. In the low current regime or \emph{subcritical} region ($ J < J_{c0} $), where $ J_{c0} $ is the zero-temperature threshold current density, finite temperature gives a nonvanishing probability for thermally activated switching. For any thermally activated process we may define an effective barrier $ U $ by fitting the mean switching rate to the form $ \tau^{-1} = \tau^{-1}_0 \exp(-U/k_BT) $. The form of the effective barrier has been shown to be approximated well by a power law,
\beq
U_{\mathrm \sigma} = U_0 (1 \mp h_{\mathrm \sigma})^{\beta} (1 \mp j),
\eeq
where $\sigma$ = AP(upper sign) or P(lower sign) labels the activation barrier for transitions out of the AP or P states, $ U_0 = (1/2)mH_{\mathrm a} $, with $ m $ the magnetic moment of the nanomagnet, and $ h_{\mathrm \sigma} = H_{\mathrm \sigma}/H_{c0} = H_{\mathrm \sigma}/H_{\mathrm a} $ and $ j = J/J_{c0} $ are dimensionless variables corresponding to the external magnetic field and current, respectively, rescaled by their zero-temperature critical values. The exponent $ \beta = 2 $ for the external magnetic field applied parallel to the easy axis and $ \beta = 3/2 $ for the external field applied 10\ensuremath{^\circ} away from the easy axis \cite{Li2004}. For a fixed value of $ j $ the switching field is determined by the attempt frequency and measurement time and thus depends on the applied magnetic field ramp-rate ($ \tau = 1$ sec), \emph{i.e.} $ \tau^{-1} \approx \tau_0^{-1} \exp(-U/k_BT) $, which for $\tau_0^{-1}=10^{-9}$ s$^{-1}$, occurs when $ U/k_BT = a \approx 20$. Using $ U = ak_BT $ for the barrier height we thus obtain the switching field
\beq
h_{\mathrm \sigma} = \pm \left\{ 1 - \left[ \frac{ak_BT}{U_0 (1 \mp j)} \right]^{1/\beta}\right\}.
\label{hS}
\eeq
Using Eq. (\ref{EffField}) we calculate the thermal-activation effective field to be
\beq
h_{\mathrm{bias}} = -\frac{1}{\beta} \left( \frac{ak_BT}{U_0} \right)^{1/\beta} \frac{j}{(1-j^2)^{1/\beta}}.
\label{hbias}
\eeq
The width of the hysteresis loop, $|h_{AP} - h_{P}|$, as a function of current is given by
\beq
h_{\mathrm{width}} = 2 \left[ 1 - \left( \frac{ak_BT}{U_0} \right)^{1/\beta} \frac{1}{(1-j^2)^{1/\beta}} \right],
\label{hwidth}
\eeq
neglecting higher order terms in $ j $ since we are interested in the regime where $ j \ll 1 $. The finite-temperature field stability boundaries determined by Eq. (\ref{hS}) are shown schematically in Figure \ref{EffFieldPD}(b). The trends are consistent with the phase diagrams shown in Figure \ref{PD}. Figures \ref{Results}(b) and (c) show the results of fitting Eqs. (\ref{hbias}) and (\ref{hwidth}), with $ J_{c0} = 1.3 \times 10^8 $ A/cm$^2$ and $ H_{c0} = H_{\mathrm a} = 320 $ Oe, to the bias field and the hysteresis loop width, respectively, in a way we detail next. In this treatment we are interested in estimating the temperature contribution to the observed STI effective field. We thus use the hysteresis loop width data, which is unchanged by the presence of any effective field, to determine the energy barrier. Fitting Eq. (\ref{hwidth}) to the hysteresis loop width as a function of current, at $ T = 295 $ K, in Fig. \ref{Results}(a), yields the value 0.67 for the fitting parameter $ A = (ak_BT/U_0)^{1/\beta} $. If we take $ \beta = 3/2 $ and $ a = 20 $ this gives an energy barrier of 0.93 eV. This is less than the value expected in a single domain picture of $ U_0 \approx 2.2 $ eV, for $ M = 1440 $ emu/cm$^3$ and $ H_{\mathrm a} = 320 $ Oe, but not unreasonable, as the magnetization reversal likely occurs non-uniformly \cite{Albert2002}. Using these parameters, we have plotted Eq. (\ref{hwidth}) for the hysteresis loop width at $ T = 4.2 $ K, in Fig. \ref{Results}(a), and Eq. (\ref{hbias}) for the effective field at $ T = 295 $ K and at $ T = 4.2 $ K, in Fig. \ref{Results}(b). The value we obtain for the energy barrier accounts for the scaling with temperature of the zero-current hysteresis loop width. Note that the measured width of the hysteresis loops decreases more rapidly than expected based on this thermal activation model, likely because the magnetization reversal processes are current dependent. For example, this current dependence could result from the circular field generated by the charge current influencing the element's domain structure. It is also clear from these curves that these fits underestimate the change in the effective field with current, both at $ T = 295 $ K and at $ T = 4.2 $ K. At $ T = 4.2 $ K the effects of thermal fluctuations are negligible near zero current, because the thermal energy is much less than the energy barrier to reversal. This is evidence for an effective field of the form $ b_J = \hat{m} \times \hat{m}_P $, \emph{in addition} to the spin-transfer torque interaction.

\section{Effective Field Interaction}
We can estimate the magnitude of a STI effective field of the form $ b_J = \hat{m} \times \hat{m}_P $ based on the measurements at $ T = 4.2 $ K. If we account for the (relatively minor) contribution of thermal fluctuations at $ T = 4.2 $ K to the current-depend shift in bias field, we obtain, near $ J = 0 $ (fit in the range, $-0.25J_{co}<J<0.25J_{co}$), $ -1.5 \times 10^{-7} $ Oe cm$^2$/A for the slope of $ b_J $. It is of interest to compare this to the magnitude of the spin-transfer torque interaction. From Eq. (\ref{eom}), the spin-transfer-torque-driven switching boundary for the AP $ \rightarrow $ P transition is given by $ a_c = \alpha(-H_{\mathrm{app}} + H_{\mathrm a} + 2 \pi M) $, while that for the P $ \rightarrow $ AP transition is given by $ a_c = -\alpha(H_{\mathrm{app}} + H_{\mathrm a} + 2 \pi M) $. If we take $ \alpha = 0.01 $, we find $ 8.0 \times 10^{-7} $ Oe cm$^2$/A, based on the measured switching current at zero applied field. To compare this quantitatively to the models \cite{Zhang2002,Shpiro} we compute the ratio of the magnitude of the spin-transfer torque $ a_J $ to that of the effective field term $ b_J $. We find $ |a_J/b_J| \approx 5.3 $.  The values in the detailed diffusive transport model of Shpiro  \emph{et al.} \cite{Shpiro} are extremely sensitive to the Co layer thickness, the interface resistance and the decay length of the transverse component of the spin accumulation, $\lambda_J$.  Within this model values of  $\lambda_J \sim 2$nm can produce a ratio close to that found in experiment. We note that the relative signs of these interactions are consistent with the models of Zhang \emph{et al.} and Shpiro \emph{et al.}

\section{Conclusions}
In summary, we find evidence for a STI effective field interaction, with a magnitude that is a factor of 5 less than the spin-transfer torque for a 3 nm thick Co free layer. Measurements of the interaction constant $b_J $ as a function of free layer thickness will be essential to understanding the origin of this interaction, such as whether it is associated with the longitudinal \cite{Heide2001b} or transverse \cite{Zhang2002,Shpiro} component of the spin accumulation. In the latter case, the interaction strength is expected to depend strongly on the free layer thickness, when this is comparable to the decay length of the transverse component of the spin accumulation, $\sim 2$ nm \cite{Zhang2002,Shpiro}. The longitudinal spin accumulation length is much longer (\emph{i.e.} 60 nm in Co) and thus these effects should thus be easy to distinguish. Future studies will also examine a wider variety of device structures and sample to sample fluctuations in these interaction constants. 

While a number experiments have been able to exclude the presence of a STI effective field without a spin-transfer torque iinteraction \cite{Myers2002,Grollier2003,Koch2004} , these results show that both interactions are likely to be present. Note that for typical nanopillar samples and comparable interaction strengths, the current induced switching threshold is still mainly determined by the spin-transfer torque interaction, while, as we have shown, at sufficiently low temperature and currents the magnetic switching threshold reflects the STI effective field interaction. The method we have presented enables a clear separation of these two mechanisms and can be applied quite generally to nanopillar devices with patterned fixed and free magnetic layers.

\begin{acknowledgments}
We thank P. M. Levy and A. Brataas for many useful discussions of this work. This research is supported by grants from NSF-FRG-DMS-0201439 and by ONR N0014-02-1-0995. M. A. Zimmler acknowledges support from the NYU Dean's Undergraduate Research Fund.
\end{acknowledgments}


\end{document}